\documentclass[journal]{IEEEtran}
\usepackage{amsfonts}
\usepackage{algorithmic}
\usepackage{algorithm}
\usepackage{array}
\usepackage{textcomp}
\usepackage{stfloats}
\usepackage{url}
\usepackage{verbatim}
\usepackage{graphicx}
\usepackage{cite}
\usepackage{mathtools, cuted}
\usepackage{lipsum}
\usepackage{subfigure} 
\usepackage{amsmath, bm} 
\usepackage{booktabs} 
\usepackage{amsthm,amssymb} 
\usepackage{dsfont} 
\usepackage{mathrsfs} 
\usepackage{color,balance} 

\begin{document}
\title{Enhanced Physical Layer Security for Full-duplex Symbiotic Radio with AN Generation and Forward Noise Suppression}
\author{Chi~Jin,~\IEEEmembership{Student Member,~IEEE,} Zheng~Chang,~\IEEEmembership{Senior Member,~IEEE,} Fengye~Hu,~\IEEEmembership{Senior Member,~IEEE,} Hsiao-Hwa Chen,~\IEEEmembership{Fellow,~IEEE,} and~Timo H\"am\"al\"ainen,~\IEEEmembership{Senior Member,~IEEE}
\thanks{C. Jin  (Email: {\tt chi.c.jin@jyu.fi}), Z. Chang  (Email: {\tt zheng.chang@jyu.fi}) and Timo H\"am\"al\"ainen  (Email: {\tt timo.t.hamalainen@jyu.fi}) are with Faculty of Information Technology, University of Jyv\"askyl\"a, FIN-40014, Jyv\"askyl\"a, Finland. F. Hu  (Email: {\tt hufy@jlu.edu.cn}) is with the  College of Communication Engineering, Jilin University, 130012, Changchun, China. Hsiao Hwa Chen (E-mail: {\tt hshwchen@ieee.org}) is with the Department of Engineering Science, National Cheng Kung University, Tainan 70101, Taiwan.}
\thanks{This work was supported by the Chinese Scholarship Council under the State Scholarship Fund (No. 202106170113), and Taiwan Ministry of Science and Technology (Nos. 109-2221-E-006-175-MY3, 109-2221-E-006-182-MY3, and 112-2221-E-006-127).}}
	
\maketitle
	
\begin{abstract}
Due to the constraints on power supply and limited encryption capability, data security based on physical layer security (PLS) techniques in backscatter communications has attracted a lot of attention. In this work, we propose to enhance PLS in a full-duplex symbiotic radio (FDSR) system with a proactive eavesdropper, which may overhear the information and interfere legitimate communications simultaneously by emitting attack signals. To deal with the eavesdroppers, we propose a security strategy based on pseudo-decoding and artificial noise (AN) injection to ensure the performance of legitimate communications through forward noise suppression. A novel AN signal generation scheme is proposed using a pseudo-decoding method, where AN signal is superimposed on data signal to safeguard the legitimate channel. The phase control in the forward noise suppression scheme and the power allocation between AN and data signals are optimized to maximize security throughput. The formulated problem can be solved via problem decomposition and alternate optimization algorithms. Simulation results demonstrate the superiority of the proposed scheme in terms of security throughput and attack mitigation performance.
\end{abstract}

\section{Introduction}
\IEEEPARstart{T}{o} embrace a human-centric, sustainable, and prosperous digital future, massive wireless devices have been deployed and the Internet of Things (IoT) applications will play an extremely important role. According to a report in the Statista \cite{2023-2030}, the number of global IoT devices reached 13 billion in 2022, and is expected to reach 30 billion in 2030. In order to support wireless connections for massive devices, spectrum/energy efficiencies and transmission security should be guaranteed \cite{IoT1,IOT-6G,iot-6g2,EURO_76GHZ,iot-spectrum}, and more effective techniques should be further developed for this purpose. 

To achieve efficient spectrum sharing and energy sustainability, symbiotic radio (SR) was proposed as a promising technique to address the spectrum sharing and energy sustainability issues \cite{sr-intro}. The proposal of SR was inspired from cognitive radio (CR) and ambient backscatter communications (AmBC) \cite{sr-intro2,sr-intro3,sr-intro4}. In an SR network, the primary transmitter (PTx) also serves as a RF signal source for a backscatter device (also called secondary transmitter (STx)). The signal backscattered from the STx to the primary receiver (PRx) contains useful information from PTx. Through a joint decoding method as mentioned in \cite{sr-model}, the STx turns out to function also as a multi-path reflector of the PTx signal rather than an interference generator, which overcomes the drawbacks of conventional CR. Thus, SR can take the advantages of both CR and AmBC, and enable PTx to power STx while communicating with PRx at the same time, improving energy efficiency and sustainability \cite{benefit1}. More importantly, It succeeds in seeking mutual benefits in a spectrum-sharing system and provides some new opportunities for designing security transmission strategies \cite{benefit2,benefit3}. 

Nevertheless, the use of STx poses a significant security challenge to a SR system, and the computational capability of the STx is one pressing limitation for enhancing transmission security. Although lightweight encryption algorithms may provide some security benefits to traditional energy-constrained sensors, they appear to be still too complex for passive backscatter devices in STx \cite{lightweight1,lightweight2,lightweight3}. Furthermore, the transmission of encryption keys remains to be exposed to eavesdroppers \cite{lightweight4}. In this context, physical layer security (PLS) is more suitable for SR. PLS ensures security transmission by maintaining the superiority of the equivalent legitimate channel over the eavesdropping channel \cite{RSMODEL}. The equivalent channels in SR can be manipulated not only at the STx (e.g., multi-user collaboration \cite{SRT-2}) but also at the PTx (e.g., artificial noise injection \cite{R01}). This means that the computational complexity at the STx can be effectively shifted to the PTx, thereby overcoming the computational bottleneck of the STx to enhance the data security.  

Among all available PLS schemes, artificial noise (AN) injection has been extensively investigated in the literatures, especially for backscatter devices (BDs). In a single-BD system, the secrecy rate was improved in \cite{R01} by optimizing AN power. A related approach for multiple antennas was proposed in \cite{R04}, and a lightweight algorithm was suggested. In \cite{R05}, the authors studied the impact of channel correlation on secrecy outage probability. When multiple BDs were available, an AN-based jamming and cooperation scheme was considered \cite{R02}. In addition, some emerging technologies, such as energy storage BDs \cite{R06} and intelligent reflecting surfaces (IRS)\cite{SR-SE01, SR-SE02}, were introduced into SR but with some challenges to address. While the aforementioned studies are indeed useful, they however considered only conventional passive eavesdropping attacks. In fact, the threat of proactive eavesdropping in SR is even more serious as it may harm the system through both direct and backscatter links \cite{R03}. Furthermore, the system optimization against proactive eavesdroppers becomes more challenging due to the complicated links and signals \cite{active-eave,active-eave2,active-eave3}. 

It is also noted that, in the earlier research efforts, two fundamental issues were typically overlooked. What motivates the primary system to provide PLS for the secondary system if they are independent? Also, the implicit assumption on the trustworthiness of the primary system in SR may introduce potential security issues for IoT devices. Specifically, to achieve joint decoding, the secondary system needs to inform the primary receiver about its backscatter symbol set. However, in practice, the secondary system cannot guarantee the trustworthiness of the primary system as they are typically independent. The situation worsens if the primary system is disguised covertly by a proactive eavesdropper, enabling effortless information theft. Fortunately, an access point (AP) centric full-duplex symbiotic radio (FDSR) can be used to avoid the above issues and more suitable to the IoT transmission paradigm \cite{FDSR-1}.

In FDSR, an AP operates in a full duplex mode, serving as a secondary receiver (SRx) and a primary transmitter (PTx) at the same time. The most prominent characteristic feature of FDSR is the primary and secondary systems are not independent \cite{FDSR-2,FDSR-3}. With respect to the SRx, AP is always pre-paired with STx, ensuring its trustworthiness. With respect to the PTx, AP can authenticate the PRx and promptly detect its abnormal behaviors. Therefore, an AP centric FDSR can avoid information leakage within the system. Another advantage of FDSR is that the AP can decode the signals from STx more effectively with less computational effort. This is because the symbols of the primary system, which impacts joint decoding performance substantially, are perfectly known to the AP. Furthermore, as the centre of the system, the AP interacts with different devices and has sufficient knowledge on the channel state information (CSI), which makes it possible to implement a PLS scheme effectively \cite{SR-CHANNEL}. 

In this paper, we aim to propose a transmission strategy to enhance the PLS of SR based on artificial noise injection, pseudo-decoding, forward noise suppression and phase control. All schemes are implemented at an AP to reduce the computational burden of STx. We want to increase the legitimate data throughput and confuse eavesdroppers, thereby improving the system security rate. Specifically, we optimize the transmission power of artificial noise and adjust the phase shift for the FDSR system with an eavesdropper, particularly a proactive eavesdropper. The main contributions of this work can be summarized as follows.
\begin{itemize}
\item 
We present a versatile FDSR transmission model with an eavesdropper that has omnidirectional/directional antennas and joint decoding capabilities. We consider a proactive eavesdropper, which can be degenerated to a passive eavesdropper by adjusting the system parameters easily. Moreover, various factors, such as self-interference cancellation and antenna gain coefficients, are taken as the parameters to leverage the system flexibility and applicability.	
\item 
We propose an artificial noise generation method, where the AN is generated, depending on the interference signal of a proactive eavesdropper and the receiver antenna noise. With the proposed pseudo-decoding and phase control techniques, the AN can help eliminate noise at PRx and improve the signal strength at STx. Furthermore, the artificial noise remains random, preserving its effectiveness to confuse eavesdroppers.
\item 
An optimization problem decomposition and alternating optimization schemes are proposed to solve the formulated problem. Insightful discussions are given to disclose a fact that the power allocation optimization problem is proved to be either boundary or extremum optimal with at most two extremum points, depending on the feasible region limited by the constraints. In addition, an analytical solution is provided for phase control as well. The results show that the proposed scheme has a low computational complexity, and its good performance is validated through extensive simulations.
\end{itemize}

The remainder of this paper is outlined as follows. Section II introduces a system model. In Section III, the optimization problem is formulated and the proposed strategy is presented. Section IV shows the numerical results with the discussions, followed by the conclusions of this paper in Section V.

\begin{table}[t]
\renewcommand{\arraystretch}{1.2}
\centering
\caption{List of notations used in this paper}
\begin{tabular}{llcc}
%\begin{tabular}{cccc}
\hline
Symbol &Definition \\ \hline
$\mathcal{A}$ &Full-duplex access point  & \\
$\mathcal{P}$ &The primary receiver \\ 
$\mathcal{S}$ &The secondary transmitter \\ 
$\mathcal{E}$ &Proactive eavesdropper \\
$h_{XY}$ &Channel coefficient from X to Y \\
$y_X$& Transmitted signal of X \\
$y_{XY}$& Received signal of Y from X\\
$x_A$ & Unit-power information signal of $\mathcal{A}$  \\
$\Gamma$ & Backscattering coefficient \\
$C$ & Equivalent unit-power signal of $\mathcal{S}$    \\
$\varphi_S$ & Phase shift caused by the circuit of $\mathcal{S}$  \\
$\varphi_A$ & Phase shift controlled by  $\mathcal{A}$  \\
$m,n$ & Power allocation factors at  $\mathcal{A}$  \\
$x_N$ &Redesigned artificial noise  \\
$z_I$ &Pseudo-information \\
$z_{PN}$ &Pseudo-noise  \\
$\theta$ &Approximation level of the
$z_I$ to the original signal\\
$\kappa $  & The equivalent antenna gain factors   \\
$\beta, \tau$ & Residual noise coefficients\\
$\alpha$ &Antenna mode index for $\mathcal{E}$  \\
$\lambda$ &Ability to decode/cancel $x_A$ at $\mathcal{E}$\\
\hline
\label{LIST OF NOTATIONS}
\end{tabular}
\end{table}
	
%\vspace{1mm}
\section{System Model and Problem Formulation}
\begin{figure}[t]
\centering
\includegraphics[height=9cm,width=9.1cm,keepaspectratio,trim=10 0 0 0, clip]{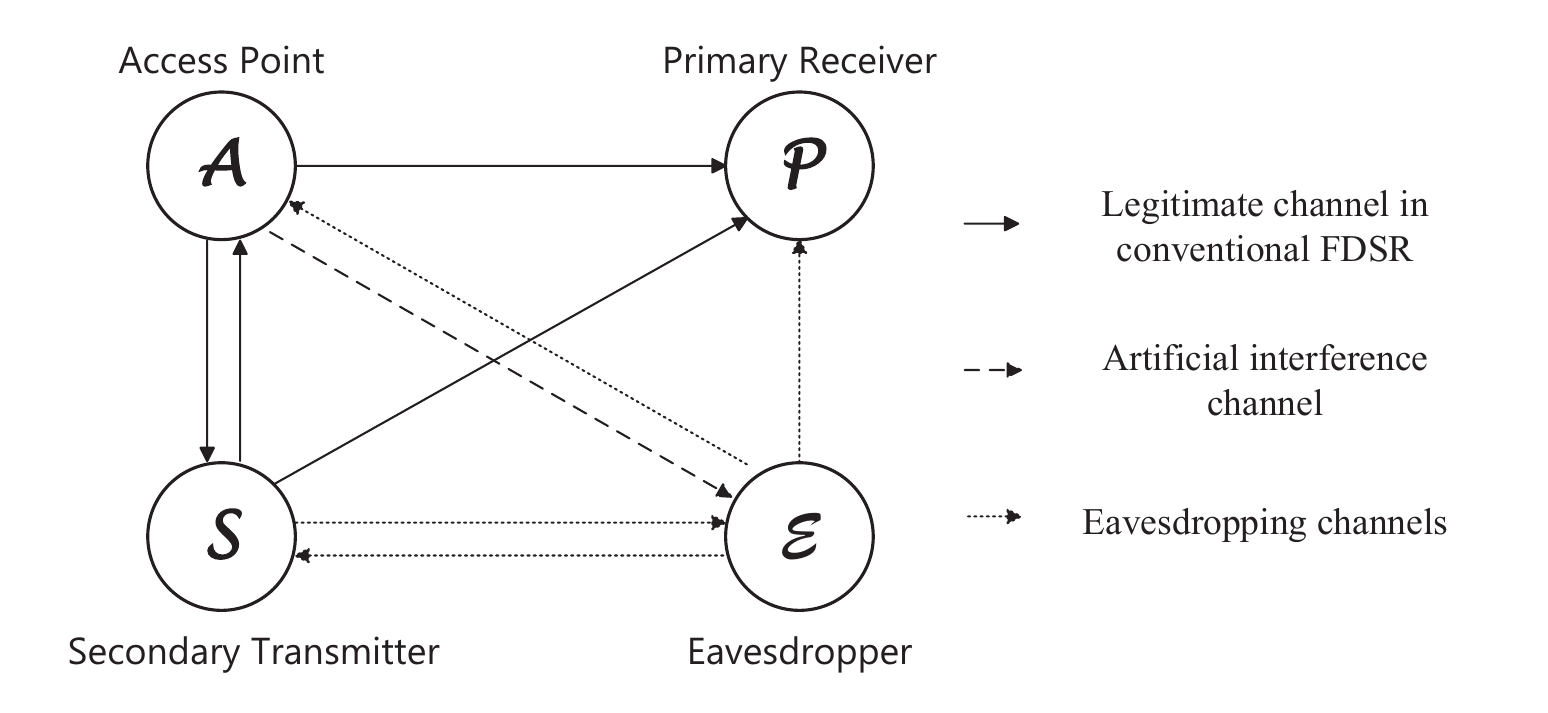}
\caption{FDSR system.}
\label{MODLE}
\end{figure}

\subsection {System Model}
Let us consider a FDSR system consisting of an AP ($\mathcal{A}$), a PRx ($\mathcal{P}$), a STx ($\mathcal{S}$) and an eavesdropper ($\mathcal{E}$) as shown in Fig. \ref{MODLE}. Specifically, $\mathcal{A}$ integrates the primary transmitter (PTx) and the secondary receiver (SRx) in a conventional SR. Therefore, $\mathcal{A}$ is responsible for sending signals to $\mathcal{P}$ as well as receiving data from $\mathcal{S}$. The FDSR system may work in two scenarios. In one scenario, the communications through $\mathcal{S}$-$\mathcal{A}$ and $\mathcal{A}$-$\mathcal{P}$ links are independent, performing spectrum sharing. In the other scenario, $\mathcal{A}$ receives the data collected by $\mathcal{S}$ and forwards it to $\mathcal{P}$, serving as a relay.
	
Assume that all devices are equipped with one receiving antenna and one transmitting antenna. $\mathcal{A}$, $\mathcal{P}$, $\mathcal{E}$ have their sufficient power supplies, but $\mathcal{S}$ transmits with the backscattered power. The backscatter coefficient of $\mathcal{S}$ is represented as $\Gamma$, and the phase shift caused by its circuit is denoted as $\varphi_S$. 
	
The channel coefficients are assumed to follow Rician fading \cite{CHANNEL1, CHANNEL2,CHANNEL03} and can be represented as
\begin{equation}
h = \left[{\sqrt{\frac{\eta}{\eta+1}}h^{\textrm {LoS}} + \sqrt{\frac{1}{\eta+1}} h^{\textrm {NLoS}}}\right] \sqrt{c_{0}\left(\frac{d}{d_{0}}\right)^{-v}},
\end{equation}
where $h^{LoS}$ denotes the line-of-sight (LoS) component with $|h^{LoS}|^2=1$. Gaussian random variable $h^{NLoS}\sim \mathcal{CN}(0, 1)$ represents the non-line-of-sight (NLoS) Rayleigh fading component. The parameter $\eta$ refers to the Rician factor, $c_0$ represents the constant attenuation due to path-loss at a reference distance $d_0$, $v$ is the path loss exponent, and $d$ corresponds to the distance between the transmitter and receiver. Channel status remains constant during each transmission time block but varies from one block to another.    
	
%\vspace{3mm}
\subsection {Attack and Defense Processes in the Primary System}
Let us consider $\mathcal{A}$ and $\mathcal{P}$ as the main system\footnote{Note the difference between the primary system consisting of PTx and PRx, and the main system here, which consists of AP $\mathcal{A}$ (which integrates PTx and SRx) and $\mathcal{P}$ (or PRx ).}. The impact of $\mathcal{E}$ on the main system is reflected in the additional interference to $\mathcal{P}$ through $\mathcal{E}$-$\mathcal{P}$ and $\mathcal{E}$-$\mathcal{S}$-$\mathcal{P}$ links, which reduces the transmission rate of the main system. To mitigate the adverse effects of $\mathcal{E}$, we combine the artificial noise method with the forward noise suppression technique and redesign the artificial noise generation strategy.
    
Assume that $\mathcal{E}$ is a proactive eavesdropper, which sends its attack signal $\hat{y}_E=\sqrt{P_E}x_E$ to the FDSR system. $x_E$ is a unit-power signal with $E\{|x_E|^2\}=1$. We assume $\mathcal{A}$ adopts the conventional artificial noise method in the secondary system, which includes $\mathcal{S}$ and SRx in $\mathcal{A}$. The transmitted signal is
\begin{equation}
\label{hat{y}_A}
{\hat{y}_A} = m{x_A} + nx^*_{N},
\end{equation}
where unit power signal $x_A$ refers to the data signal, $x^*_{N}$ is the artificial noise, $m$ and $n$ represent power allocation factors. Thus, we can derive 
\begin{equation}
m^2 + n^2\sigma^2_N = P_A,
\end{equation}
where $P_A$ is the transmission power budget of $\mathcal{A}$. In this case, the signals received at $\mathcal{P}$ from other devices are
\begin{equation}
{y_{AP}}={h_{AP}}{e^{ - jw{t_{AP}}}} \left( {m{x_A} + nx^*_{N}} \right),
\end{equation}
\begin{equation}
{y_{EP}}={h_{EP}}{e^{ - jw{t_{EP}}}} \sqrt{P_E}x_E ,
\end{equation}
\begin{equation}
{y_{SP}}={h_{SP}}{e^{( - jw{t_{SP}}+\varphi_S)}} \Gamma C\left( y_{AS}+y_{ES} \right),
\end{equation}
where $\Gamma$ represents the backscattering coefficient, $C$ is the equivalent unit-power baseband signal of $\mathcal{S}$, and $\varphi_S$ is the phase shift caused by the circuit. $y_{AS}$ and $y_{ES}$ are given as
\begin{equation}
{y_{AS}}={h_{AS}}{e^{ - jw{t_{AS}}}} \left( {m{x_A} + nx^*_{N}} \right),
\end{equation}
\begin{equation}
{y_{ES}}={h_{ES}}{e^{ - jw{t_{ES}}}}\sqrt{P_E}x_E .
\end{equation}
	
Based on the above derivations, the received signal at $\mathcal{P}$ is
\begin{equation}
\label{y_p1}
{{y}_{P}}= y_{AP}+y_{EP}+y_{SP}+n_P,
\end{equation}
in which ${{y}_{P}}$ consists of four components, i.e., $x_A$, $x_E$, $x^*_N$, and antenna noise $n_P\sim(0,\sigma^2_P)$. To mitigate the impact of $\mathcal{E}$ with forward noise suppression and phase control, the artificial noise $x^*_{N}$  can be redesigned as
\begin{equation}
\label{AN}
{x}_{N}= y_{EA}+n_A,
\end{equation}
where $y_{EA}$ is $\mathcal{A}$'s received signal from $\mathcal{E}$, which is
\begin{equation}
{y_{EA}}= {h_{EA}}{e^{ - jw{t_{EA}}}}  \sqrt {{P_E}} {x_E}.
\end{equation}
Therefore, the transmitted signal of $\mathcal{A}$ can be designed as 
\begin{equation}
\label{hat_y_A_pri}
{\hat{y}_A} = m{x_A} + ne^{j\varphi_{A}}(y_{EA}+n_A),
\end{equation}
where $\varphi_{A}$ is the phase shift control at $\mathcal{A}$, $y_{EA}$ is used for forward noise suppression, and $n_A\sim(0,\sigma^2_P)$ represents the antenna noise of $\mathcal{A}$ with known statistical properties. Since $n_A$ is random, the artificial noise item ${x}_{N}=y_{EA}+n_A$ is random too. Based on this, if a passive eavesdropper is present, the artificial noise depends only on the antenna noise (i.e., ${x}_{N}=n_A$), and then our strategy is equivalent to the conventional AN injection scheme.
	
With the revised transmitted signal (\ref{hat_y_A_pri}), the received signal at $\mathcal{P}$ is rewritten as
\begin{equation}
\begin{split}
\label{y_p2}
{y}_{P} &= {h_{AP}}{e^{ - jw{t_{AP}}}} \left\{ m{x_A} + e^{j\varphi_{A}}n(y_{EA}+n_A) \right\}  \\
&\quad +{h_{AS}}{h_{SP}}{e^{ - (jw{t_{SP}}+jw{t_{AS}}-\varphi_S)}}  \Gamma C \\ 
&\quad \times \left\{ m{x_A} + e^{j\varphi_{A}}n(y_{EA}+n_A) \right\} \\
&\quad +{h_{ES}}{h_{SP}}{e^{- (jw{t_{SP}}+jw{t_{ES}}-\varphi_S)}} \Gamma C  \sqrt{P_E}x_E  \\
&\quad +{h_{EP}}{e^{ - jw{t_{EP}}}} \sqrt{P_E}x_E +n_P.
\end{split}
\end{equation}
As the backscattered signal carries the information from $\mathcal{A}$, the $\mathcal{P}$ can treat the $\mathcal{A}$-$\mathcal{S}$-$\mathcal{P}$ link as an extra NLoS link through the joint decoding strategy to improve data throughput. By adjusting the power and phase of ${x}_{N}$ (i.e., $m$ and $\varphi_A$), forward noise suppression for $\mathcal{A}$ is available.

\begin{figure}[t]
\centering
\includegraphics[height=7.3cm,width=7.3cm,keepaspectratio,trim=0 0 0 0, clip]{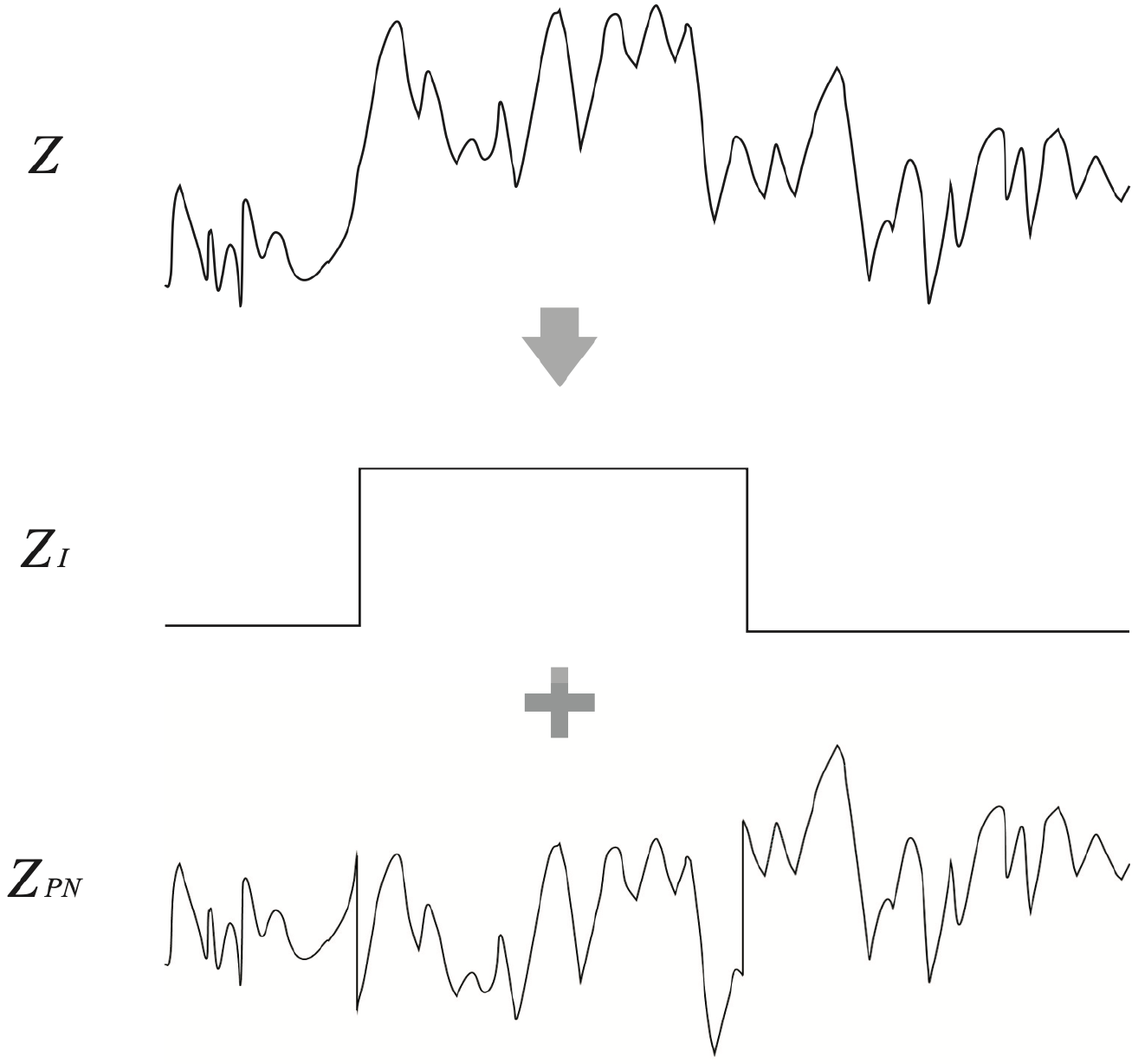}
\caption{Pseudo-decoding process for signal $z$.}
\label{PD}
\end{figure}

%\vspace{3mm}
\subsection {Attack and Defense Process in Secondary System}
The secondary system suffers eavesdropping over the $\mathcal{E}$-$\mathcal{S}$-$\mathcal{E}$ link and interference through the $\mathcal{E}$-$\mathcal{A}$ and $\mathcal{E}$-$\mathcal{S}$-$\mathcal{A}$ links. We focus on enhancing the security transmission rates of PLS here, and will briefly discuss about the ways to combine advanced interference suppression techniques in the next subsection. 

Let us define the security transmission throughput of the secondary system as 
\begin{equation}
\label{RS}
_S = max\left\{0, R_A - R_E\right\},
\end{equation} 
where $R_A$ and $R_E$ represent the signal receiving rates of $\mathcal{A}$ and $\mathcal{E}$\cite{RSMODEL}. 
    
The theoretical basis of the artificial noise injection method is to establish the superiority of an equivalent legitimate channel over an equivalent eavesdropping channel artificially. Without artificial noise, the received signal at $\mathcal{E}$ can be represented as 
\begin{align}
\hat{y}^{*}_{E} &=\Gamma C h_{SE}{e^{ - jw{t_{SE}}}}
h_{AS}{e^{ - jw{t_{AS}}}}\sqrt{P_A}x_A \nonumber \\
&\quad+\Gamma C h_{SE}{e^{ - jw{t_{SE}}}} h_{ES}{e^{ - jw{t_{ES}}}}\sqrt{P_E}x_E \nonumber \\
&\quad+ h_{AE}{e^{ - jw{t_{AE}}}}\sqrt{P_A}x_A+n_E.
\end{align}
Since the statistical properties of the eavesdropper $\mathcal{E}$ are unknown, in extreme cases where $x_A$ is perfectly decoded and $\sigma^2_E \rightarrow 0$, the eavesdropping rate may potentially become exceptionally high. To solve this problem, artificial noise injection is adopted, and the received signal can be written as 
\begin{align}
\hat{y}^{}_{E} &=\Gamma C h_{SE}{e^{ - jw{t_{SE}}}}h_{AS}{e^{ - jw{t_{AS}}}}(mx_A+nx_N) \nonumber \\
&\quad+\Gamma C h_{SE}{e^{ - jw{t_{SE}}}}h_{ES}{e^{ - jw{t_{ES}}}}\sqrt{P_E}x_E \nonumber \\
&\quad+ h_{AE}{e^{ - jw{t_{AE}}}}(mx_A+nx_N)+n_E.
\end{align}
Since $\mathcal{E}$ lacks prior information about $x_N$, it cannot eliminate the artificial noise component. However, with perfect knowledge of artificial noise, $\mathcal{A}$ can remove the artificial noise. This asymmetry of information flows establishes the superiority of the equivalent legitimate channel. 

In brief, artificial noise injection interferes the eavesdropping channel by consuming part of the transmit power to create a throughput gap $R_S$. To further enhance the system performance, we propose the pseudo-decoding scheme as follows. 
	
A pseudo-decoding process means to use partial or incomplete information to approximate the original signal. The pseudo-decoding does not care about the decoded data and can be performed at any bit rate, and thus its result may not be unique. As shown in Fig. \ref{PD}, signal $z$ is decomposed into pseudo-information $z_{I}$ ($01100$) and pseudo-noise $z_{PN}$ through pseudo-decoding.

When the first item of artificial noise in Eqn. (\ref{AN}) is pseudo-decoded, $x_N$ can be decomposed into
\begin{equation}
\label{x_n_pd}
x_N=z_{I}+z_{PN}+n_A=z_I+z_N,
\end{equation}
where we have
\begin{align}
&\sigma^2_N=h^2_{EA}P_E+\sigma^2_A,\\
&\sigma^2_I=\theta h^2_{EA}P_E,\\
&\sigma^2_{PN}=(1-\theta) h^2_{EA}P_E,
\end{align}
where $\theta$ refers to the approximation level of the pseudo-information to the original signal, which depends on the modulation order and the bit rate of the pseudo-decoding scheme. The most ideal state is $\theta\rightarrow1$, which means most of the artificial noise power can be converted into effective signal power by pseudo-decoding.
	
When adjusting $z_{I}$ to the same bit time as $x_A$, the mask encryption is realized. Furthermore, $x_A$ and $z_{I}$ can be considered as plaintext and mask, with the mask known only to $\mathcal{A}$. When analyzing the secondary system, the signal sent from $\mathcal{A}$ can be rewritten as
\begin{equation}
\label{hat_y_A_sec}
{\hat{y}_A} = m{x_A} + e^{j\varphi_{A}}n(z_I+z_N)=x_{CW}+ne^{j\varphi_{A}}z_N,
\end{equation}
where $m{x_A} + ne^{j\varphi_{A}}z_I$ is regarded as the mask-encrypted carrier wave (CW) for backscatter communications. Therefore, the mask can also be viewed as a valid signal by $\mathcal{A}$. The received signal of $\mathcal{A}$ is
\begin{align}
\begin{split}
\label{y_a}
{y}_{A} &= h_{AS}h_{SA}{e^{-({ jw{t_{AS}}+jw{t_{SA}}-\varphi_s})} }\left( x_{CW}+ne^{j\varphi_{A}}z_N \right)\\
&+h_{ES}h_{SA}{e^{-({ jw{t_{ES}}+jw{t_{SA}}-\varphi_s})} }\sqrt{P_E}x_A\\
&+{h_{EA}}{e^{ - jw{t_{EA}}}}  \sqrt{P_E}x_E.
\end{split}
\end{align}
	
On the other hand, the eavesdropper $\mathcal{E}$ receives signals from three links, i.e., $\mathcal{A}$-$\mathcal{E}$, $\mathcal{A}$-$\mathcal{S}$-$\mathcal{E}$, and $\mathcal{E}$-$\mathcal{S}$-$\mathcal{E}$ links, resulting in a received signal denoted as
\begin{align}
\label{y_e}
{y}_E &= h_{SE}e^{\varphi_S} \Gamma C h_{ES}\sqrt{P_E}x_E \nonumber \\
&\quad+ h_{SE}e^{\varphi_S} \Gamma C  h_{AS}(x_{CW}+ne^{j\varphi_{A}}z_N) \nonumber \\
&\quad+ h_{AE} \left( x_{CW} + ne^{j\varphi_{A}}z_N \right),
\end{align}
from which we can see that the waveform of random AN signal does not affect the interference to $\mathcal{E}$ when its power remains unchanged. Therefore, the use of pseudo-decoding can improve the throughput of the secondary system, while confusing $\mathcal{E}$ effectively.

\newpage
\begin{strip}
\begin{equation}
\label{T2}
T^2={{({\kappa _2}{h_{SP}} \Gamma C {h_{AS}})}^2} + {{({\kappa _3}  {h_{AP}})}^2} + 2{\kappa _2}  {\kappa _3}  {h_{SP}}  \Gamma C  {h_{AS}}{h_{AP}}\cos ({{\varphi_{SP}} - {\varphi _S} + {\varphi_{AS}} - {\varphi_{AP}}}).
\end{equation}
\begin{equation}
\label{U2}
U^2={P_E}\left[ {{{({\kappa _1}  {h_{EP}})}^2} + {{({\kappa _2}  {h_{SP}}  \Gamma C  {h_{ES}})}^2} + 2{\kappa _1}  {\kappa _2}  {h_{EP}}{h_{SP}}  \Gamma C  {h_{ES}}\cos ({\varphi_{EP}} -  {{\varphi_{SP}} -{\varphi_{ES}} +{\varphi _S}} )} \right].
\end{equation}
\begin{equation}
\label{V2}
V^2= { {{h^2_{EA}}}}{P_E}\left[ {{{({\kappa _2}  {h_{SP}}  \Gamma C  {h_{AS}})}^2} + {{({\kappa _3}  {h_{AP}})}^2} + 2{\kappa _2}  {\kappa _3}  {h_{SP}}  \Gamma C  {h_{AS}}{h_{AP}}\cos ({\varphi_{SP}} - {\varphi _S} + {\varphi_{AS}} - {\varphi_{AP}})} \right].
\end{equation}
\begin{equation}
\label{fai1}
\begin{split}
\varphi_1&=\arctan \frac{{{\kappa _1}  {h_{EP}}\sin ({\varphi_{EP}}) + {\kappa _2}  {h_{SP}}  \Gamma C  {h_{ES}}\sin ({\varphi_{SP}} + {\varphi_{ES}} - {\varphi _S})}}{{{\kappa _1}  {h_{EP}}\cos ({\varphi_{EP}}) + {\kappa _2}  {h_{SP}}  \Gamma C  {h_{ES}}\cos ({\varphi_{SP}} + {\varphi_{ES}} - {\varphi _S})}} \\
&- \arctan \frac{{{\kappa _2}  {h_{SP}}  \Gamma C  {h_{AS}}\sin ({\varphi_{SP}} - {\varphi _S} + {\varphi_{AS}} + {\varphi_{EA}}) + {\kappa _3}  {h_{AP}}\sin ({\varphi_{AP}} + {\varphi_{EA}})}}{{{\kappa _2}  {h_{SP}}  \Gamma C  {h_{AS}}\cos ({\varphi_{SP}} - {\varphi _S} + {\varphi_{AS}} + {\varphi_{EA}}){\rm{ }} + {\rm{ }}{\kappa _3}  {h_{AP}}\cos ({\varphi_{AP}} + {\varphi_{EA}})}}.
\end{split}
\end{equation}
\begin{equation}
\label{J2}
J^2=\sigma _A^2\left[ {{{({\kappa _3}  {h_{AP}})}^2} + {{({\kappa _2}  {h_{SP}}  \Gamma C  {h_{AS}})}^2}} \right].
\end{equation}
\begin{equation}
\label{HAT_M}
\hat{M}={ {P_A}\theta{{\left( {{h_{SA}}{h_{AS}}\Gamma C} \right)}^2}}.
\end{equation}
\begin{equation}
\label{HAT_Q}
\hat{Q}= (1 - \theta ){{\left( {{h_{SA}}{h_{AS}}\Gamma C} \right)}^2}.
\end{equation}
\begin{equation}
\label{L}
\begin{split}
L&=\tau{{{P_E}\left[ {{{({h_{SA}}  \Gamma C  {h_{ES}})}^2} + {{({h_{EA}})}^2} + 2{h_{SA}}  \Gamma C  {h_{ES}}{h_{EA}}\cos ({\varphi_{SA}} + {\varphi_{ES}} - {\varphi _S} + {\varphi_{EA}})} \right] }}\\&+{\beta P_A(1 - \theta ){{\left( {{h_{SA}}{h_{AS}}\Gamma C} \right)}^2}}+ \sigma _A^2.
\end{split}
\end{equation}
\begin{equation}
\label{R}
R={-\beta (1 - \theta ){{\left( {{h_{SA}}{h_{AS}}\Gamma C} \right)}^2}}.
\end{equation}
\begin{equation}
\label{HAT_A}
\hat{A}={{\left( {{h_{SE}}\Gamma C h_{ES}} \right)}^2} {P_E}.
\end{equation}
\begin{equation}
\label{HAT_B}
\hat{B}={{\left( \alpha\lambda{{h_{AS}}\Gamma C h_{SE}} \right)}^2}.
\end{equation}
\begin{equation}
\label{C}
C={{P_A} \left[ {{{\left( {{h_{SE}}\Gamma C  {h_{AS}}} \right)}^2} + {{\left( {\alpha   {h_{AE}}} \right)}^2} + 2{h_{SE}}\Gamma C  {h_{AS}}\alpha   {h_{AE}}\cos (w{t_{SE}} - {\varphi _S} + w{t_{AS}} - w{t_{AE}})} \right]  + \sigma _E^2}.
\end{equation}
\begin{equation}
\label{D}
D={ - {\lambda ^2}\left[ {{{\left( {{h_{SE}}\Gamma C  {h_{AS}}} \right)}^2} + {{\left( {\alpha   {h_{AE}}} \right)}^2} + 2{h_{SE}}\Gamma C  {h_{AS}}\alpha   {h_{AE}}\cos (w{t_{SE}} - {\varphi _S} + w{t_{AS}} - w{t_{AE}})} \right]}.
\end{equation}
\hrule
\end{strip}

\subsection { Security Transmission Problem }
Eqns. (\ref{hat_y_A_pri}) and (\ref{hat_y_A_sec}) are different representations of $\mathcal{A}$'s transmitted signal, because forward noise suppression and pseudo-decoding are considered simultaneously when designing the artificial noise signal. 
 
Next, we will add some additional parameters to the model to extend it to more scenarios. First, let us consider a combination scheme with multi-antenna technology. Although we assume that all devices are equipped with only a single transmitting antenna and a single receiving antenna, the impact of multi-antenna technology on system performance can be investigated by introducing an equivalent antenna gain.

Let us define the equivalent antenna gains between $\mathcal{P}$ and $\mathcal{E}$, $\mathcal{P}$ and $\mathcal{S}$, $\mathcal{P}$ and $\mathcal{A}$ as $\kappa_1$, $\kappa_2$ and $\kappa_3$. According to (\ref{y_p2}), the signal-to-noise ratio (SNR) at $\mathcal{P}$ can be written as
\begin{equation}
\begin{split}
\label{gamma_p}
{\gamma_P} = \frac{{{m^2}   {T^2}}}{{\left( {{U^2} + {G^2}} \right) + {n^2}\left( {{V^2} + {J^2}} \right) - 2nUV\cos ({\varphi _A} + {\varphi _1})}},
\end{split}
\end{equation}
where $G^2=\sigma _P^2$, $T^2$, $U^2$, $V^2$, $J^2$ and $\varphi_1$ are given in (\ref{T2}) $\sim$ (\ref{J2}).
	
Second, assume that $\mathcal{A}$ is able to eliminate interference from $z_N$ and $x_E$, and denote the residual noise coefficients as $\beta$ and $\tau$, where $\beta, \tau \in [0,1]$. If $\beta$ takes the value of 0, it represents that the noise $z_N$ is completely cancelled, and this is true for $\tau$ and $x_E$.  Then, based on (\ref{y_a}),  the SNR at $\mathcal{A}$ can be written as
\begin{equation}
\label{GAMMA_A}
{\gamma_A} = \frac{\hat{M}+\hat{Q}m^2}{L+Rm^2},
\end{equation}
where $\hat{M}$, $\hat{Q}$, $L$ and $R$ are given in (\ref{HAT_M}) $\sim$ (\ref{R}).
	
Finally, since the properties of the eavesdropper $\mathcal{E}$ are always unknown, we make the three assumptions to evaluate our strategy explicitly as follows.
\begin{enumerate}
\item[a)]
We use $\alpha$ to represent the eavesdropper's antenna mode, where $\alpha = 1$ indicates an omnidirectional antenna  and $\alpha = 0$ means a directional antenna.
\item[b)]
We assume that eavesdropper's ability to decode $x_A$ (when $\alpha = 1$)  or to cancel $x_A$ (when $\alpha = 0$) is denoted by the factor $\lambda$. When $\lambda = 0$, $x_A$ is purely an interference for $\mathcal{E}$. The $\mathcal{E}$ received signal $x_A$ through both the $\mathcal{A}$-$\mathcal{E}$ and $\mathcal{A}$-$\mathcal{S}$-$\mathcal{E}$ links, and the eavesdropping rate is suppressed by an increasing interference power. However, when $\lambda = 1$, $x_A$ is perfectly cancelled or decoded. Thus, the $x_A$ component from the $\mathcal{A}$-$\mathcal{E}$ link can be eliminated. Moreover, the $\mathcal{A}$-$\mathcal{S}$-$\mathcal{E}$ link may help to improve eavesdropping rate under the omnidirectional antenna scheme through joint decoding.  
\item[c)]
We consider a scenario, where $\mathcal{E}$ achieves an optimal performance, i.e., $\lambda = 1$ and $\sigma_E^2 = 0$. In this case, the security transmission rate  $R_S$ approaches the lower bound. 
\end{enumerate}
 
Based on the aforementioned assumptions and (\ref{y_e}), the SNR of the eavesdropper can be expressed as
\begin{equation}
\label{GAMMA_E}
{\gamma_E} = \frac{\hat{A}+\hat{B}m^2}{C+Dm^2},
\end{equation}
where $\hat{A}$, $\hat{B}$, $C$ and $D$ are given in (\ref{HAT_A}) $\sim$ (\ref{D}).

For analysis simplicity, let $M = \hat{M} + L$, $Q = \hat{Q} + R$, $A = \hat{A} + C$, and $B = \hat{B} + D$. Then, we can formulate an optimization problem as follows.
\begin{equation}
\label{P1}
\begin{split}
\hspace{-1.9mm}&(P1):\quad\quad{\max \limits_{m, \ \varphi_A}\ R_S},\\
&s.t.\quad  
\begin{array}{lc}
C1:  R_P\ge R^{th}_P,\\
C2: m^2 + n^2 \sigma^2_N = P_A,\\
C3: m\ge 0, n\ge 0.
\end{array}
\end{split}
\end{equation}

Our objective is to maximize the security rate of the secondary system, i.e., $R_S$, with the following constraints. The throughput of the primary system is guaranteed and the optimization variable $\varphi_A$ is reflected in $C1$. $C2$ represents the power constraint for data signal and artificial noise, and $C3$ restricts the feasible ranges of power allocation factors $m$ and $n$.

%\vspace{3mm}
\section{Proposed Approach}
Substituting (\ref{GAMMA_A}) and (\ref{GAMMA_E}) into (\ref{P1}) and applying the properties and monotonicity of the logarithmic function, we can rewrite $R_S$, or the objective function of ($P1$), as
\begin{equation}
\hat{R_S}=\frac{M+Qm^2}{L+Rm^2} \frac{C+Dm^2}{{A}+{B}m^2}.
\end{equation}
To solve ($P1$), an alternating optimization method is adopted, where in each iteration one variable is fixed, and the other variable is optimized alternately until the convergence condition is satisfied.
	
\subsection {Optimum AN Power Allocation}
Let us first consider $\varphi_A$ as a constant. Then, the optimization problem ($P1$) can be transformed to
\begin{equation}
\label{P2}
\begin{split}
\hspace{-1.9mm}(P2):\quad\quad&\max \limits_{m}\ \frac{M+Qm^2}{L+Rm^2} \frac{C+Dm^2}{{A}+{B}m^2},\\
&\hspace{-16mm}s.t.\quad  
\begin{array}{lc}
\vspace{3mm}
C1: \dfrac{{ m^2   {T^2}}}{{\left( {{U^2} + {G^2}} \right) + {n^2}\left( {{V^2} + {J^2}} \right) - 2nUV\phi}} \ge \gamma^{th}_P,
\\
C2: m^2 + n^2(h^2_{EA}P_E+\sigma^2_A) = P_A,
\vspace{1mm}\\
C3: m\ge 0, n\ge 0,
\end{array}
\end{split}
\end{equation}
where $\gamma^{th}_P=2^{R^{th}_P}-1$, and $\phi=\cos ({\varphi _A} + {\varphi _1})$. Based on $C2$ and $C3$, we can solve constraint $C1$ and obtain the feasible regions for the power allocation factors $m$ or $n$. Thus, ($P2$) can be solved by searching the optimum values of $m$ or $n$ within the feasible regions in
\begin{equation}
\label{NR}
\begin{split}
\hspace{-2mm}\bigl\{n\ &\big| \ n \in \mathbb{R},\ {n^2} \left[ {{r_P}\left( {{V^2} + {J^2}} \right) + (h_{EA}^2{P_E} + \sigma _A^2){T^2}} \right]  \\&- 2n {{r_P}UV {\phi} }   +  {{r_P}\left( {{U^2} + {G^2}} \right) - {P_A}{T^2}}  \le 0 \bigr\}.
\end{split}
\end{equation}
Correspondingly, the feasible region of $m$ can be written as $m \in [m_{c1}, m_{c2}]$.
\vspace{2mm}
 
Denote the objective function of ($P2$) as $f(m)$, whose trend can be determined by examining its first derivative
\begin{equation}
\label{FOD}
\frac{{\partial f(m)}}{{\partial m}} = \frac{a m^2+b m+c}{\{(L+Rm^2)(A+Bm^2)\}^2}=0,
\end{equation}
where we set
\begin{align}
&a=(AD-BC)QR+(BD+D^2)(LQ-MR),\\
%\end{equation}
%\begin{equation}
&b=2\left[(A+C)DLQ-(B+D)CMR\right],\\
%\end{equation}
%\begin{equation}
&c=AC(LQ-MR)+(AD-BC)LM+C^2(LQ-MR).
\end{align}
 
Since the denominator of Eqn. (\ref{FOD}) is the product of the interference power of $\mathcal{A}$ and the received power at $\mathcal{E}$, which should not become zero, we can transform Eqn. (\ref{FOD}) into a quadratic equation as
\begin{equation}
\label{FODT}
a m^2+b m+c=0.
\end{equation}
Now, we can solve Eqn. (\ref{FODT}) and obtain two solutions $m_1$ and $m_2$ ($m_1\leq m_2$), which are the extreme points of the objective function $f(m)$, and also the candidates for optimal $m$ in ($P2$).

\vspace{2mm}
\emph{\textbf{Theorem $1$}}: 
Define $m_{1}, m_{2}$ as $m'_{1}, m'_{2}$ if $m_{1}, m_{2}\in [m_{c1}, m_{c2}]$. The optimum power allocation factor $m^*$ can be obtained simply by searching from the given set $\{m_{c1}, m_{c2}, m'_1, m'_2\}$. Also, the analytical solutions are given in the Appendix.

\vspace{2mm}
%\begin{proof}
\emph{Proof}. Please refer to Appendix A. \hfill\IEEEQEDclosed
%\end{proof}

\subsection {Optimum Phase Control}
To investigate the impact of the phase control factor $\varphi_A$ on ($P1$), let us assume $m$ is a constant and use Lagrangian multiplier method to transform ($P1$) into an optimization problem with $\phi=\cos(\varphi_A+\varphi_1)$ and $\mu$ as its variables, or
\begin{equation}
\label{P3}
\begin{split}
\hspace{-1mm}(P3):\quad&\max \limits_{\phi,\ \mu}\ f(m)\\&+ \mu\left\{\frac{{ m^2   {T^2}}}{{\left( {{U^2} + {G^2}} \right) + {n^2}\left( {{V^2} + {J^2}} \right) - 2nUV\phi}}-\gamma^{th}_P \right\} \\
&s.t.\quad  
\begin{array}{lc}
C1: -1\leq \phi \leq 1\\
C2: \mu \geq 0
\end{array}
\end{split}
\end{equation}
Next, we establish a new optimization problem as
\begin{equation}
\label{P4'}
\begin{split}
\hspace{-1mm}(P3'):\quad&\min \limits_{\phi}\ \left({{U^2} + {G^2}} \right) + {n^2}\left( {{V^2} + {J^2}} \right) - 2nUV\phi\\
&s.t.\quad  
\begin{array}{lc}
-1\leq \phi \leq 1\\
\end{array}
\end{split}
\end{equation}

\vspace{2mm}
\emph{\textbf{Theorem $2$}}: 
The optimization problem ($P3'$) is equivalent to ($P3$). Thus, the solution of ($P3$) can be achieved by solving the single-variate optimization problem ($P3'$).

\vspace{2mm}
\emph{Proof}. Please refer to Appendix B. \hfill\IEEEQEDclosed

\vspace{2mm}
Then, the objective function can be rewritten as
\begin{equation}
\label{FF}
\begin{split}
f(\phi)&=\left({ {G^2}+n^2{J^2} } \right) + \left( U^2+n^2V^2-2nUV\phi \right) \\
&=f_1+f_2(\phi).
\end{split}
\end{equation}
Due to the constraint $-1\leq \phi \leq 1$, we have
\begin{equation}
\label{EQ2}
(U-nV)^2\leq f_2(\phi)\leq (U+nV)^2.
\end{equation}
Therefore, if the phase $\varphi_A$ is continuously adjustable, $f_2(\phi)$ reaches its minimum value when $\phi^*=1$, that is 
\begin{equation}
\label{PHIC}
\varphi^*_A=-\varphi_1,
\end{equation}
which is also the optimum phase control parameter for ($P3$). 
 
On the other hand, if the phase control is discrete, the two closest discrete values to the continuous optimal value $-\varphi_1$ should be used.

\vspace{2mm}
\emph{\textbf{Theorem $3$}}: 
According to (\ref{EQ2}), it can be shown that the optimum discrete phase control is
\begin{equation}
\label{PHID}
\varphi^*_A\in \{(-\varphi_1)^+, (-\varphi_1)^-\},
\end{equation}
where $(\varphi)^+$ and $(\varphi)^-$ indicate the nearest available discrete phases around $\varphi$.

\vspace{2mm}
\emph{Proof}. Please refer to Appendix C. \hfill\IEEEQEDclosed

\begin{table}[t]
\normalsize
\renewcommand{\arraystretch}{1.0}
\centering
\begin{tabular}{l}
\toprule
\textbf{Algorithm 1} Joint alternating optimal solution\\
\midrule 
1:  \textbf{Input:}  $\bm h$, $\bm P$, $\Gamma$, and $\delta$.\\
2:  \textbf{Output:} $m^*$, $\varphi^*_A$;\\
3:	\textbf{for} $itera = 1$ to $20$, \textbf{do}\\
4:   ~~~~Setting $\phi=1$ and obtain $m^*$ according \\
~~~~~~~~to \emph{{Theorem $1$}};\\
5:  ~~~~Obtain $\varphi_A$ according to (\ref{PHIC});\\
6:~~~~~\textbf{If} $\varphi^*_A$ is continuously adjustable: \\
7:~~~~~~~ Jump to step 9; \\
8: ~~~~~~~\textbf{else if} $\varphi^*_A$ is discretely adjustable: \\
9:~~~~~~~~~~~ Calculate discrete $\varphi^*_A$ based on \emph{{Theorem $3$}};\\
10:~~~~~~~~~~ Update the feasible region of $m$ according \\
~~~~~~~~~~~~~~~to (\ref{NR});\\
11:~~~~\textbf{end if}\\
12:~~~~Calculate security throughput $\mathcal{T}_{itera}$ with (\ref{RS});\\
13: ~~~\textbf{If} ($\mathcal{T}_{itera}-\mathcal{T}_{itera-1}\le$ $\delta$)\\
13:	~~~~~~~Return $m^*$, $\varphi^*_A$ and terminate;\\
14:~~~~\textbf{end if}\\
15:\textbf{end for}\\
\bottomrule
\end{tabular}
\end{table}

%\vspace{2mm}
\subsection {Joint Optimization}
Based on the above derivations, we propose a joint alternating optimization algorithm as shown in Algorithm 1, which can be summarized as optimizing and updating variables alternately until the convergence of the objective function. In practice, $\Gamma$ is determined by the circuit, $\bm h$  can be estimated or modeled, and $\delta$ is the convergence threshold.
	
If the phase $\varphi_A$ is continuously adjustable, the algorithm can always obtain an optimal solution because setting $\phi=1$ will always be feasible. In this case, the two sub-problems are completely decoupled, and the problem is degenerated into optimizing $m$ under the condition $\phi=1$. For the discrete phase optimization, a high-quality sub-optimal transmission strategy $\{m, \varphi_A\}$ is available for the proposed FDSR system.

The proposed algorithm has a low computational complexity. Specifically, the complexity is dominated by (\ref{NR}), (\ref{FODT}) and (\ref{fai1}), all of which have a complexity of $\mathcal{O}(1)$. Therefore, the algorithm is of $\mathcal{O}(k)$ complexity, where $k$ is the iterations.

\section{Simulation Results}
In this section, we conduct several numerical results to verify the effectiveness of the proposed strategy and to evaluate its performance. Unless otherwise specified, we assume the noise variance $\sigma^2_A = \sigma^2_E=\sigma^2_P= -80$ dBm, and the reflection coefficient $\Gamma= 0.7$ for all devices \cite{RC0.6, RC0.9}.  The channel-related parameters are path loss exponent $v = 3$, the Rician factor $\eta=3$, and the constant attenuation $c_0=-20$ dB\cite{CHANNEL2}. The transmission power and eavesdropper power budgets are $P_{Amax}=P_{Emax} = 2$ W, and they both transmit at the maximum power\cite{POWER,POWER2}. The decoding or noise cancel factor of $\mathcal{E}$ is set to $\lambda=1$ to evaluate the lower bound of system performance. The throughput constraint of the primary system is $\gamma^{th}_P=10$, and $\mathcal{P}$ uses single receiver antenna, that is, $\kappa_1=\kappa_2=\kappa_3=1$. To simulate the actual spatial location, the distances between all devices are randomly generated with uniform distributions $d \sim U(1,8)$ m \cite{DISTANCE, DISTANCE2}, and the channel phase shifts $\varphi \sim U(0, 2\pi)$\cite{PHASE}.
	
In the following subsections, the performance of the proposed strategy (PS) and the widely studied AN scheme \cite{R04,R05,R01,R02} in FDSR systems are compared, with the security throughput as the evaluation metric. In the simulations, both cases that  $\mathcal{E}$ is equipped with the omnidirectional antenna (OA) and the directional antenna (DA) are considered. Likewise, insights on additional system parameters such as throughput constraint $\gamma^{th}_P$, the reflection coefficient $\Gamma$, and the decoding factor of the eavesdropper $\lambda$ are also provided for more flexible scheme design.
	
To simplify the description, $P_A$, $P_E$, and the power of data signal at $\mathcal{A}$ denote transmit power, interference power, and signal power in the following analysis.

\begin{figure*}[t]
\centering
\hspace{-6mm}
\subfigure[Security throughput]{
\includegraphics[height=9cm,width=9.3cm,keepaspectratio,trim=10 0 0 0, clip]{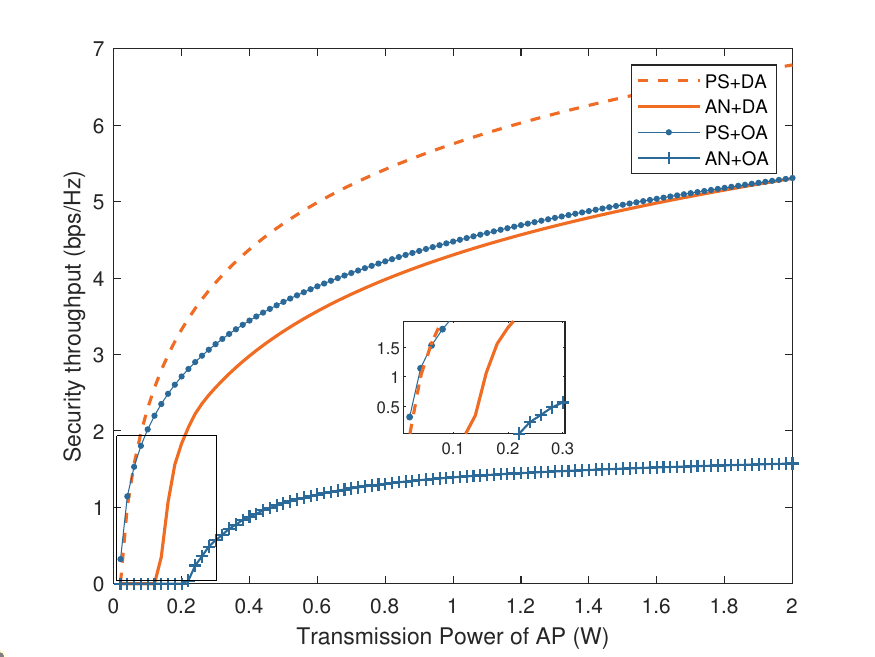}}
\hspace{-6mm}
\centering
\subfigure[ Optimum signal power and DAR]{
\includegraphics[height=9cm,width=9.3cm,keepaspectratio,trim=10 0 0 0, clip]{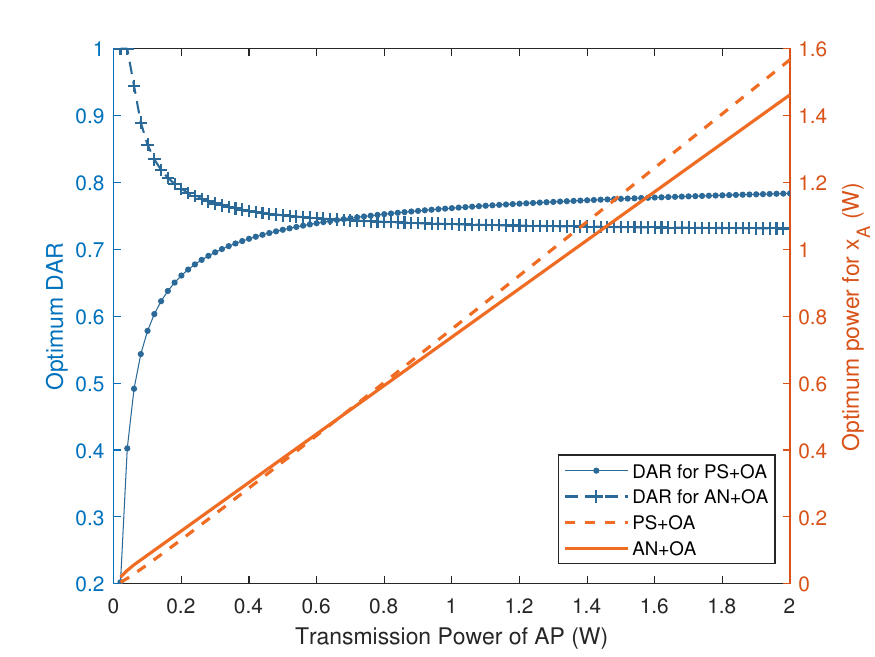}}
\caption{Security throughput, optimum signal power, and DAR versus the transmission power $P_A$.}
\label{FIG PA}
\end{figure*}

\begin{figure*}[t]
\centering
\hspace{-6mm}
\subfigure[Security throughput]{
\includegraphics[height=9cm,width=9.3cm,keepaspectratio,trim=10 0 0 0, clip]{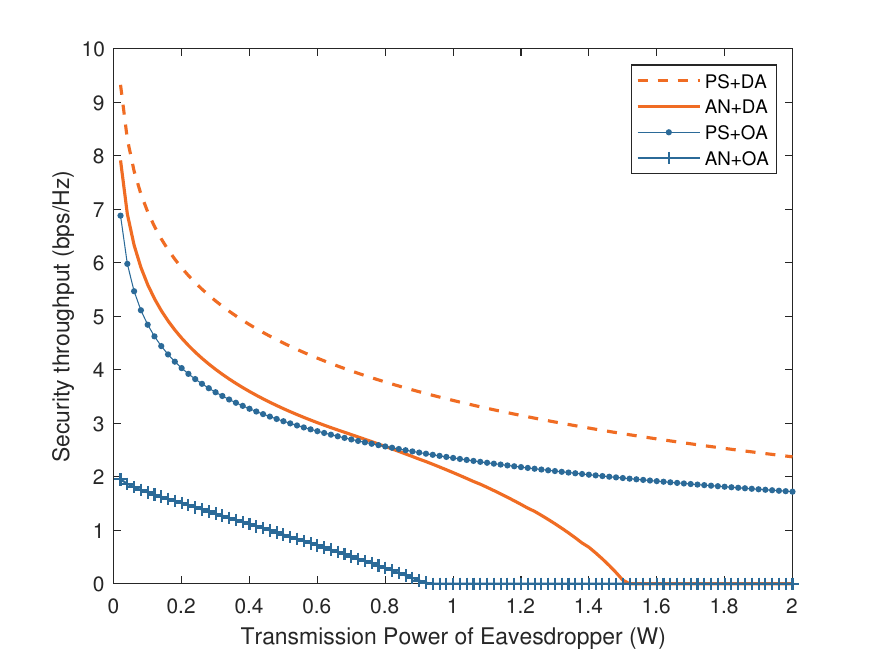}}
\hspace{-6mm}
\centering
\subfigure[Optimum signal power and DER]{
\includegraphics[height=9cm,width=9.3cm,keepaspectratio,trim=10 0 0 0, clip]{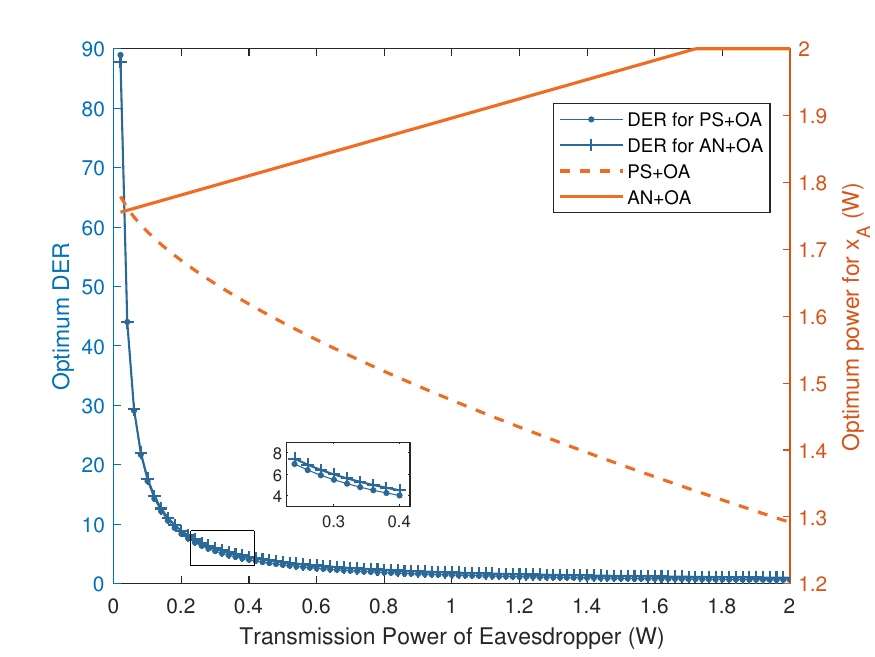}}
\caption{Security throughput, optimum signal power, and DER versus the transmission power $P_E$.}
\label{FIG PE}
\end{figure*}

\subsection {Impacts of Transmission Power $P_A$}
Fig. \ref{FIG PA}(a) shows how the security throughput performance varies with the transmit power with different transmission strategies. Here, `PS+DA' means the PS scheme is adopted at $\mathcal{A}$, and $\mathcal{E}$ is equipped with a directional antenna (DA). Basically, the PS scheme outperforms the AN scheme in both OA and DA cases. The security throughput increases with all strategies as $P_A$ increases.
	
In the zoomed sub-figure, it is clearly illustrated that the minimum operating powers for the strategies are different, which are influenced by the primary system's throughput constraint $\gamma^{th}_P$. The PS schemes have lower power requirements with the redesigned AN and the forward interference suppression technique. Furthermore, comparing 'PS+DA' and 'PS+OA' strategies, we can see that $\mathcal{E}$ with  DA does not always result in a lower security throughput. This is because DA can cancel the interference from $\mathcal{A}$-$\mathcal{E}$ link, but also eliminate the possibility of jointly decoding the data signal $x_A$, which may help to increase the wiretap channel throughput in $\mathcal{A}$-$\mathcal{S}$-$\mathcal{E}$ link. To evaluate the lower bound of the throughput performance, we set the decoding factor as $\lambda=1$ in the simulations, which means that $\mathcal{E}$ can perfectly decode or cancel $x_A$. Therefore, in most value ranges of $P_A$, the OA scheme achieves a worse performance than DA.
	
Fig. \ref{FIG PA}(b) shows the optimal signal power allocation and the corresponding DAR under different transmit powers $P_A$, where DAR is defined as data signal power to transmit power ratio. An interesting result is that the optimal signal powers allocated for the PS and AN schemes are generally similar, but their DARs show an opposite trend. The reason is that under a lower power budget, the primary task of the AN scheme is to satisfy the constraint, and thus more power is allocated to $x_A$. For PS, due to the phase control, forward noise suppression, and the pseudo-decoding method, the components of the data signal and the AN signal are optimized. As a result, constraints can be satisfied easily, making a smaller DAR feasible. With an increasing $P_A$, both DARs tend to stabilize.

In summary, the security throughput increases with $P_A$, and the proposed PS scheme can operate with a lower transmit power and consistently achieve a higher security throughput.

\subsection {Impacts of Transmission Power $P_E$}
Fig. \ref{FIG PE}(a) shows the effect of the interference power ($P_E$) on the security throughput. Intuitively, as $P_E$ increases, the security throughput decreases gradually. Compared to the AN scheme, the PS scheme has a better performance in the whole range. 

Besides, it can be observed that the PS scheme is more tolerant to a higher $P_E$. With an increasing $P_E$, the throughput decreasing in the PS scheme shows a slowing trend, while throughput in the AN scheme drops to 0 quickly after a brief slowdown. The reason is that when $P_E$ is small, the increase of $P_E$ obviously affects the SNR at $\mathcal{A}$ and $\mathcal{P}$. When $P_E$ is larger, subject to the primary system throughput constraint $\gamma^{th}_P$, the feasible range of the power allocation factor $m$ keeps shrinking. At this time, the performance of the AN scheme experiences a second rapid decline. Taking the advantages of the forward interference suppression and phase control, the PS scheme can still meet the constraint at a higher $P_E$, overcoming the drawbacks of the AN scheme.
	
Similarly, the optimal power allocation factor $m$ and DER of the OA scheme, which is defined as the data signal power to interference power ratio, is considered in Fig. \ref{FIG PE}(b). Different from Fig. \ref{FIG PA}(b), here the DERs of PS and AN schemes are similar, while the optimal signal power shows a reverse trend. With the increase of $P_E$, the AN scheme needs a higher signal power to satisfy the constraint. Under the premise to satisfy the constraints, PS avoids the joint-decoding $x_A$ at $\mathcal{E}$ by reducing signal power, and uses pseudo-decoding to ensure the transmission rate between $\mathcal{A}$ and $\mathcal{S}$, thereby reducing the impact of $P_E$ on security throughput.
	
To summarize, as $P_E$ increases, the security throughput drops, and the proposed PS scheme can operate effectively even in the presence of a higher interference power, consistently achieving a superior security throughput.

\begin{figure}[t]
\centering
\includegraphics[height=9cm,width=9.3cm,keepaspectratio,trim=20 0 0 0, clip]{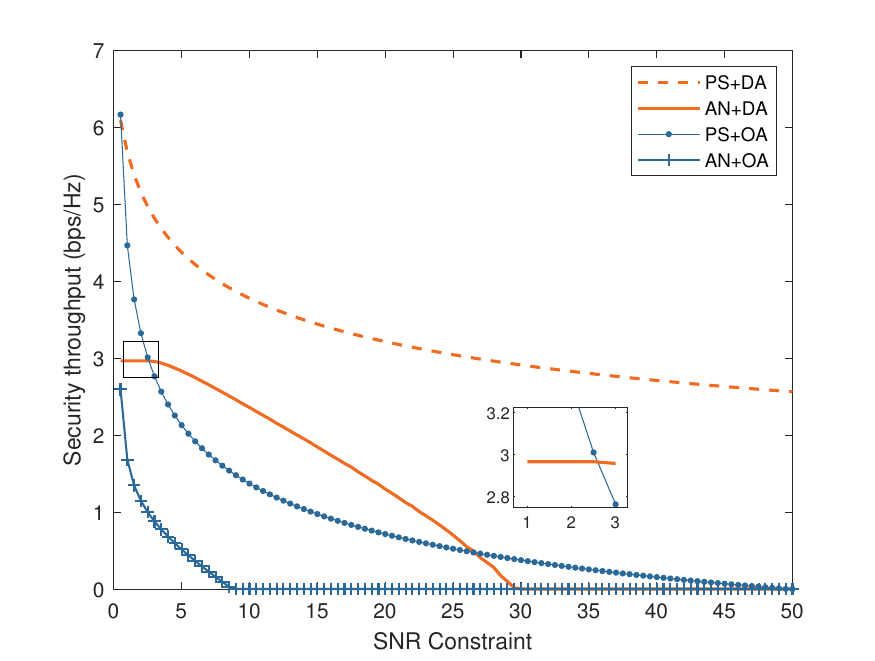}
\caption{Security throughput versus the SNR constraint $\gamma^{th}_P$.}
\label{FIG CONS}
\end{figure}
	
\begin{figure}[t]
\centering
\includegraphics[height=9cm,width=9.3cm,keepaspectratio,trim=20 0 0 0, clip]{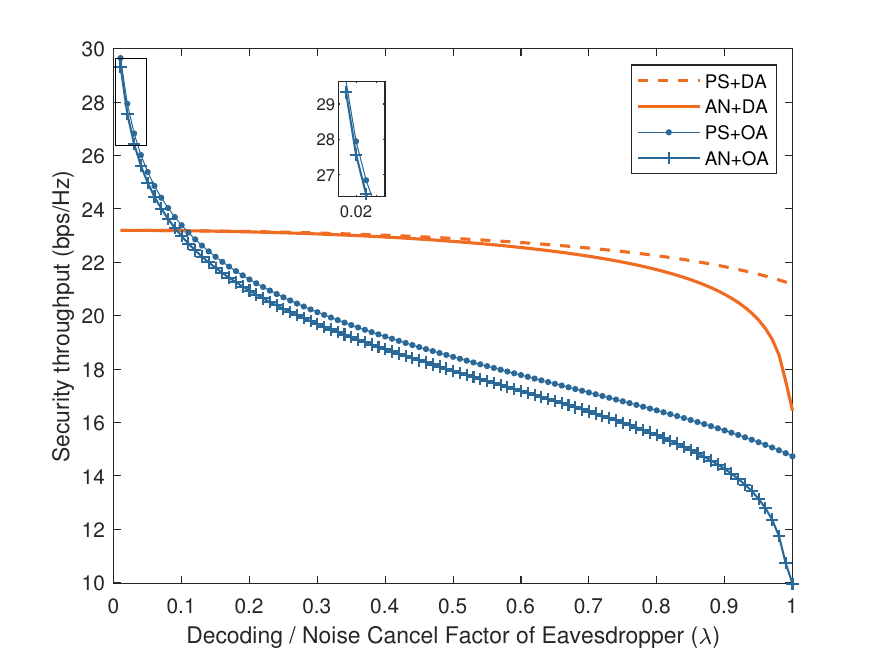}
\caption{Security throughput versus the decoding or noise cancel factor $\lambda$. }
\label{FIG DECODE}
\end{figure}

\subsection {Impacts of Constraint $\gamma^{th}_P$}
Fig. \ref{FIG CONS} reflects that the security throughput decreases as the primary system throughput constraints ($\gamma^{th}_P$) increase. The PS scheme is still usable and consistently achieves a higher throughput with larger SNR constraints. The principle that $\gamma^{th}_P$ affects the throughput is to change the feasible region $m\in[m_{c1},m_{c2}]$ and the extreme points $\{m_1,m_2\}$ of the power allocation factor $m$, thereby affecting the numerical relationship between them. According to (\ref{M1}) and (\ref{M2}), $m^*$ is influenced by the numerical relationship and further affects the security throughput.
	
Note that, as shown in the zoomed sub-figure, when the constraint takes a small value, the throughput may remain constant within a certain interval. This usually happens when the extreme point lies in the feasible region. At this time, the shrinkage of the feasible region does not affect the optimal value of $m$. When the feasible region is further reduced, $m^*$ is no longer included and the throughput begins to decline, as the `AN+DA' strategy shown in Fig. \ref{FIG CONS}.

\subsection {Impacts of Decoding / Noise-cancel Factor $\lambda$}
The decoding or noise-cancel factor $\lambda$ is the most special parameter in the proposed strategy. In the OA and DA schemes, $\lambda$ has different meanings. Under the DA scheme, $\lambda$ represents the ability of $\mathcal{E}$ to eliminate the data signal $x_A$ backscattered by $\mathcal{S}$, where $x_A$ is viewed as noise. Whereas under the OA scheme, $\mathcal{E}$ receives $x_A$ signal from the $\mathcal{A}$-$\mathcal{E}$ link and may be able to joint-decode $x_A$ and the date of $\mathcal{S}$. Thus, it is critical to explore the impact of $\lambda$, especially the tolerance of different strategies to it.
	
Fig. \ref{FIG DECODE} shows the security throughput as a function of $\lambda$. An important observation is that as $\lambda$ increases, the throughput decreases and the advantage of the PS over the AN scheme becomes more evident. When $\lambda=0$, that is, when $\mathcal{E}$ does not have decoding or noise cancellation capabilities, the main advantage of the PS scheme over AN is the increase in the transmission rate between $\mathcal{S}$ and $\mathcal{A}$. With the enhancement of $\mathcal{E}$'s capability, the superiority of the PS scheme is gradually highlighted. In practice, the capability of $\mathcal{E}$ is usually unknown, and thus the PS scheme is more reliable in guaranteeing security throughput due to its tolerance to high $\lambda$. 
	
In addition, it can be observed that the OA scheme is more sensitive to the rise of $\lambda$. In the OA scheme, the $\mathcal{A}$-$\mathcal{E}$ link brings in the noise to $\mathcal{E}$ and provides the possibility of joint decoding. A high $\lambda$ results in little noise interference and a huge increase in eavesdropping throughput, which leads to a rapid descent in the performance of the AN scheme.
	
All in all, the increase of $\lambda$ leads to a decrease in security throughput, PS is superior under high $\lambda$, and the OA scheme is affected more by $\lambda$.
	
\begin{figure}[t]
\centering
\includegraphics[height=9cm,width=9.1cm,keepaspectratio,trim=20 5 10 5, clip]{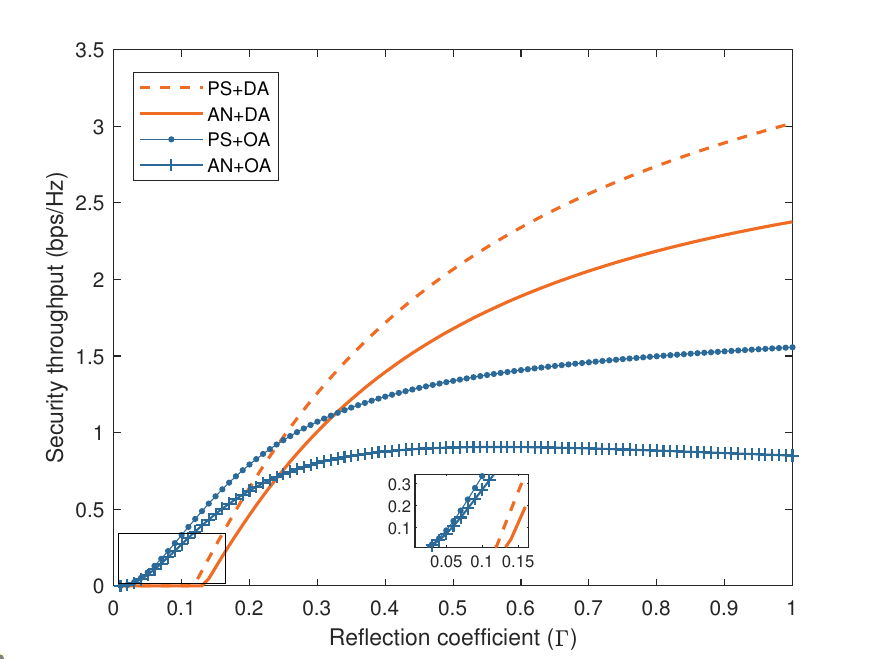}
\caption{Security throughput versus the reflection coefficient $\Gamma$.}
\label{FIG GAMMA}
\end{figure}

\subsection {Impacts of Reflection Coefficient $\Gamma$}
The reflection coefficient is the most overlooked parameter in backscatter communications. Intuitively, the larger $\Gamma$ is, the better the system performance will be. However, for the SR system with proactive $\mathcal{E}$, this is not always true.
	
As shown in Fig. \ref{FIG GAMMA}, with the increase of $\Gamma$, the security throughput of each strategy increases, but the throughput of the `AN+OA' strategy begins to decrease when $\Gamma>0.45$. The reason is that $\mathcal{S}$ not only returns data through the $\mathcal{A}$-$\mathcal{S}$-$\mathcal{A}$ link and injects artificial noise through the $\mathcal{A}$-$\mathcal{S}$-$\mathcal{E}$ link, but also helps $\mathcal{E}$ eavesdrop through the $\mathcal{E}$-$\mathcal{S}$-$\mathcal{E}$ link and interferes with $\mathcal{A}$ through the $\mathcal{E}$-$\mathcal{S}$-$\mathcal{A}$ link. Therefore, in the conventional OA scheme, the impact of $\Gamma$ on security throughput depends on the result of the game between the above two players. Excitingly, the proposed PS scheme can avoid performance degradation at high $\Gamma$ largely
	
Note that when $\Gamma$ is small, the security throughput may be zero, which means that under the given $\Gamma$, the legitimate throughput is smaller than the wiretap throughput. This situation is more likely to occur in DA schemes because of the lower noise power at $\mathcal{E}$. However, as $\Gamma$ increases, the throughput of DA rises faster and the performance gap between DA and OA scheme increases gradually.
	
Overall, both the security throughput of the PS scheme and the performance gap with the AN scheme increase with $\Gamma$. The DA scheme needs a larger lower bound of $\Gamma$ but benefits more obviously from the increase of $\Gamma$.

\begin{figure}[t]
\centering
\includegraphics[height=9cm,width=9.1cm,keepaspectratio,trim=20 5 10 5, clip]{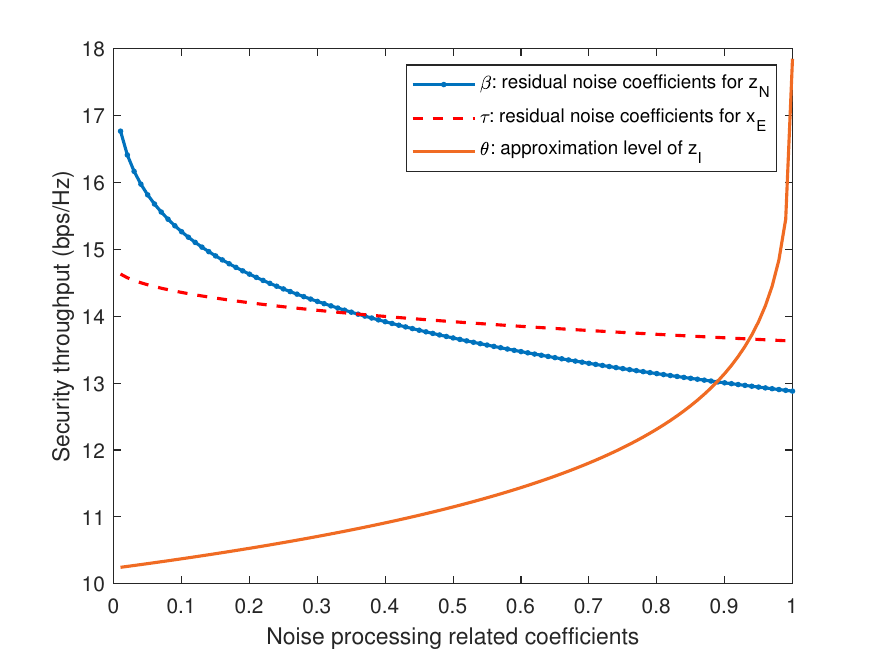}
\caption{Security throughput versus the noise-processing related coefficients}
\label{FIG 3V}
\end{figure}

\subsection{Impact of Noise-processing Related Coefficients $\beta$, $\tau$, $\theta$}
Fig. \ref{FIG 3V} shows the impact of the noise-processing-related coefficients on the security throughput performance in the proposed scheme. Generally, better noise processing requires more computing resources, and Fig. \ref{FIG 3V} provides a guidance on the trade-off between resources and performance.

Under the 'PS+OA' strategy, we studied residual noise coefficients for $z_N$ ($\beta)$, residual noise coefficients for $x_E$ ($\tau$), and approximation level of $z_I$ ($\theta$). It can be observed that the throughput decreases with $\beta$ and $\tau$ because the increase in residual noise directly affects the SNR. On the other hand, security throughput increases with $\theta$, and its impact is usually more obvious because increasing $\theta$ increases data signal power and reduces interference simultaneously.  As mentioned in Section II, $\theta$ can be improved by adjusting the modulation order and the bit rate of the pseudo-decoding scheme. Since pseudo-decoding does not concern about pseudo-information as well as symbol rate, it is easier to implement and more flexible.

Therefore, compared to noise cancellation ($\beta$ and $\tau$), improving pseudo-decoding performance ($\theta$) is more effective in terms of resource cost and performance.

\begin{figure*}[t]
\centering
\hspace{-6mm}
\subfigure[Security throughput under different discrete phase schemes.]{\includegraphics[height=9cm,width=9.1cm,keepaspectratio,trim=10 0 0 0, clip]{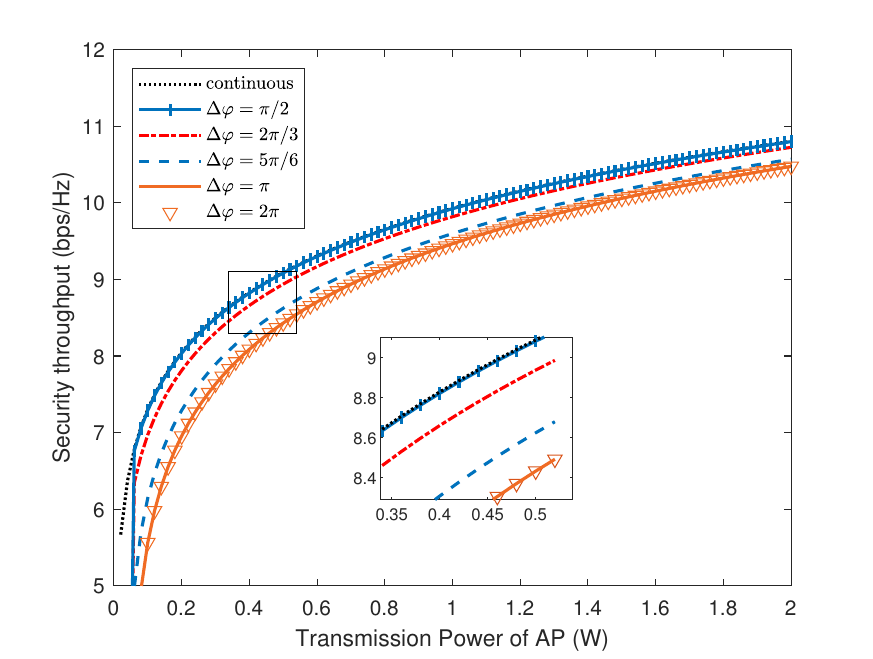}}
\hspace{0mm}
\centering
\subfigure[Security throughput versus phase.]{\includegraphics[height=9.0cm,width=9.1cm,keepaspectratio,trim=10 0 0 0, clip]{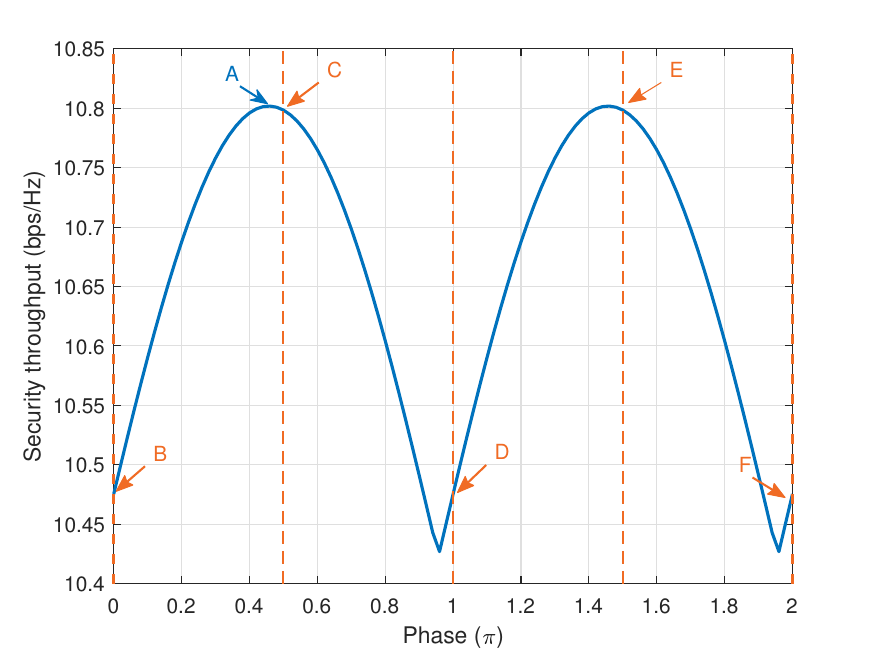}}
\caption{Comparison of continuous and discrete phase optimization schemes.}
\label{FIG Phi}
\end{figure*}

%\vspace{3mm}
    
\subsection{Comparison of Continuous and Discrete Phase Optimization Schemes}
Fig. \ref{FIG Phi} shows a comparison of continuous and discrete phase optimization schemes. In the discrete phase optimization, the phase is restricted to discrete values. This means that only a limited set of phases is available for optimization, rather than the entire continuous phase space. 
    
Fig. \ref{FIG Phi}(a) shows the secure throughput under different discrete phase schemes, where $\Delta \varphi_A$ is the phase resolution. Taking $\Delta \varphi_A=\pi/2$ as an example, the feasible phase set is $\hat{\varphi}\in\{0, \pi/2, \pi, 3\pi/2\}$ (where $2\pi$ is equivalent to zero, indicating no phase optimization). It can be observed that with the increase of $\Delta \varphi_A$, the performance of secure throughput keeps on decreasing, but not linearly. Under the given parameter settings, when $\Delta \varphi_A=\pi/2$, the performance is only slightly degraded, and the most prominent impact is that the discrete phase optimization may not work properly at a low transmit power. When $\Delta \varphi_A=\pi$, its performance is exactly the same as when $\Delta \varphi_A=2\pi$ (no phase optimization). To further explain this phenomenon, we plot the security throughput as a function of phase, as shown in Fig. \ref{FIG Phi}(b).

In Fig. \ref{FIG Phi}(b), the blue line shows the security throughput versus the phase under $P_A=2$. It can be observed that the throughput is a periodic function with period $\pi$. The reason is that the optimal value of $m$ is determined by $[m1, m2, mc1, mc2]$ in joint optimization, and $\cos^2(\varphi_A)$ appears in the process of solving these variables. According to the trigonometric identity theorem, we can get $\cos^2(\varphi_A)=[\cos(2\varphi_A)+1]/2$, leading the period on $\varphi_A$ as $\pi$. 

Three different discrete schemes are considered in Fig. \ref{FIG Phi}(b), i.e., $\Delta \varphi_A=\pi/2$, $\pi$ and $2\pi$. When $\Delta \varphi_A=\pi/2$, the throughputs corresponding to the available values of $\varphi$ are marked by points $B$, $C$, $D$, $E$, and $F$. As shown in the figure, the optimal solution is $\varphi_A=\pi/2$ (i.e., point $C$). Similarly, point $B$, $D$, and $F$ are feasible when $\Delta\varphi_A=\pi$. Note that points $B$, $D$, and $F$ are equivalent due to the periodicity, which means that $\varphi_A$ is not optimized. This explains why the curves representing the two cases in Fig.\ref{FIG Phi}(a) coincide. For the other two discrete phase schemes ($\Delta \varphi_A=2\pi/3$ and $5\pi/6$) in Fig. \ref{FIG Phi}(a), the throughput performance can be found within the interval ($0.5\pi,\pi$) correspondingly. Another noteworthy phenomenon is under the given parameters, the optimal phase for strategy $\Delta\phi_A=\pi/2$ and continuous phase optimization are very close, which explains why their curves almost overlap in Fig. \ref{FIG Phi}(a). Further, we can conclude that to achieve the optimum performance under a certain parameter setting, the minimum $\Delta\varphi_A$ depends on the optimal continuous phase $\varphi^*_A=-\varphi_1$. Since $\varphi_1$ is randomly distributed, the smaller $\Delta\varphi_A$ is, the closer the performance of discrete phase optimization is to the continuous.

In general, compared to continuous phase optimization, discrete phases will reduce the security throughput performance. The more discrete phases available in the interval $(0,2\pi)$, the greater the security throughput will be.

%\vspace{5mm}
\section{Conclusion}
	
In this paper, we studied the security transmission issue in the FDSR systems. Security throughput was improved through a novel artificial noise generation and phase control strategy. First, we introduced a transmission paradigm of the FDSR system. Second, we focused on the security transmission strategy based on artificial noise and forward noise suppression and proposed a pseudo-decoding method to improve the effectiveness of artificial noise. In particular, we verified the superiority of the proposed scheme in multiple modes and analyzed the impact of different parameters. Third, we obtained the optimal solution with a low computational complexity through problem decomposition and joint optimization. Numerical results showed that the security throughput increases with the transmit power budget ($P_A$), and decreases with the attack power ($P_E$) and the main system throughput constraints ($\gamma^{th}_P$). In addition, compared to the conventional AN strategies, the proposed strategy is more robust in hostile communication environments. 

\appendices
%\vspace{5mm}
\section{Proof of the Proposition 1}
In order to find the optimum power allocation factor $m^*$, we first need to determine its feasible region. Subject to the primary system throughput constraint, the feasible region of variable $n$ is given in (\ref{NR}). According to $m^2 + n^2 \sigma^2_N = P_A,$ we get the corresponding feasible region for $m$, that is $m\in (m_{c1}, m_{c2})$.

Next, we need to calculate the extreme points for the objective function of ($P2$), which depends on Eqn. (\ref{FODT}). When Eqn. (\ref{FODT}) has no real solutions, $f(m)$ is monotonous and the optimal solution lies on the boundary points of the feasible region, i.e., $m^*\in \{m_{c1}, m_{c2}\}$. Otherwise, $m^*\in \{m_{c1}, m_{c2}, m_1, m_2\}$ and some more specific discussions are required. 

According to the values of $a$, $m_{c1}$, $m_{c2}$, $m_1$ and $m_2$, the optimal solution of ($P2$) can be classified into four cases as follows.
	
{\textit{Case 1}}: $a>0$ and $m_1,m_2 \in \mathbb{I}$. The solutions are imaginary numbers, so the first derivative of $f(m)$ is always greater than zero, indicating that $f(m)$ is monotonically increasing with $m$. In this case, the optimal solution is the upper bound of $m$, that is $m^* = m_{c2}$.
	
{\textit{Case 2}}: $a\leq 0$ and $m_1,m_2 \in \mathbb{I}$. The first derivative is always negative and $f(m)$ is a monotonically decreasing function. Thus, it holds that $m^* = m_{c1}$. 
	
{\textit{Case 3}}: $a> 0$ and $m_1,m_2 \in \mathbb{R}$. The first derivative is in ``U" shape and has two zero points. In this case, the trend of $f(m)$ is "rising-falling-rising". The optimum $m$ is determined by the relationship between $m_{c1}$, $m_{c2}$, $m_1$, and $m_2$. Finally, it stands that
\begin{equation}
\label{M1}
m^*=\left\{
\begin{aligned}
&m_{c2}, &m_{c1} < m_{c2} < m_1 \ or\  m_{c1}>m_2,\\
&m_1, & m_{c1} < m_1 < m_{c2} < m_2,\\
&F\left\{m_1, m_{c2}\right\}, & m_{c1} < m_1 < m_2 < m_{c2},\\
&m_{c1},& m_1 < m_{c1} < m_{c2} < m_2,\\
&F{\{m_{c1}, m_{c2}\}}, &  m_1 < m_{c1} < m_2 < m_{c2},\\
\end{aligned}
\right.
\end{equation}
where $F\{x_1, x_2\}$ means $argmax_{\left\{x_1, x_{2}\right\}}\ f(m)$.
	
{\textit{Case 4}}: $a\leq 0$ and $m_1,m_2 \in \mathbb{R}$. The first derivative is in ``$\cap$" shape. Thus, the trend of $f(m)$ is ``falling-rising-rising", and it can be asserted that
\begin{equation}
\label{M2}
m^*=\left\{
\begin{aligned}
&m_{c1}, &m_{c1} < m_{c2} < m_1 \ or\  m_{c1}>m_2,\\
&F{\{m_{c1}, m_{c2}\}}, & m_{c1} < m_1 < m_{c2} < m_2,\\
&F\left\{m_{c1}, m_{2}\right\}, & m_{c1} < m_1 < m_2 < m_{c2},\\
&m_{c2},& m_1 < m_{c1} < m_{c2} < m_2,\\
&m_2, &  m_1 < m_{c1} < m_2 < m_{c2}.\\
\end{aligned}
\right.
\end{equation}
	
To summarize, the optimum power allocation factor $m^*$ can be obtained through the above analytical solutions. Furthermore, we describe $m_{1}, m_{2}$ as $m'_{1}, m'_{2}$ if they lie in the feasible region of $m$. Then, $m^*$ can be achieved simply by searching within the given set $\{m_{c1}, m_{c2}, m'_{1}, m'_{2}\}$. \hfill\IEEEQEDclosed

\section{Proof of the Proposition 2}
Let us show the equivalence of optimization problems ($P3$) and ($P3'$) by proving that they have the same KKT conditions. Since the constraint $-1\leq \phi \leq 1$ is common to problems ($P3$) and ($P3'$), we omit it in the following discussions, without affecting the proof of equivalence.

For ($P3$), the first derivative of its Lagrangian function is given as
\begin{equation}
\label{LP3}
\nabla L_1= \frac{{ 2nUV\mu m^2  {T^2}}}{{\left[\left( {{U^2} + {G^2}} \right) + {n^2}\left( {{V^2} + {J^2}} \right) - 2nUV\phi\right]^2}}. \\
\end{equation}
For ($P3'$), the first derivative of its Lagrangian function is
\begin{equation}
\label{LP33}
\nabla L_2= -2nUV. \\
\end{equation}
As discussed in Section III, the denominator of $\Delta L_1$ is always positive, so the Stationarity  Conditions in KKT can be simplified as
\begin{equation}
\label{LP30}
\nabla L_1= 2nUV\mu=0, \\
\end{equation}
\begin{equation}
\label{LP330}
\nabla L_2= -2nUV=0. \\
\end{equation}

For ($P3$), when $\mu$ takes the value of zero, the objective function is a constant. Thus, $\mu=0$ is meaningless, and the constraint of ($P3$) degenerates to $\mu>0$. At this time, Problems ($P3$) and ($P3'$) have the same Stability Conditions as
\begin{equation}
\label{LL}
\nabla L_1= \nabla L_2=2nUV=0. \\
\end{equation}
Therefore, the equivalence of Problems ($P3$) and ($P3'$) has been proved. \hfill\IEEEQEDclosed
 
\section{Proof of the Proposition 3}
According to (\ref{PHIC}), the optimum continuous phase $\varphi^*_A=-\varphi_1$. Due to the periodicity of the objective function, we discuss about discrete phase optimization in the interval $\varphi^d_A=[-\varphi_1, -\varphi_1+2\pi]$. Let $\varphi^d=\varphi_1+\varphi^d_A$. Then, we get $\varphi^d \in [0,2\pi]$. It is noteworthy that $\cos(\varphi^d)$ is monotonically decreasing in the interval $[0,\pi]$, monotonically increasing in the interval $[\pi,2\pi]$, and symmetric about $\varphi^d=\pi$. Therefore, we assume that the feasible discrete phase set is $\mathbf{\varphi}^D=\{\varphi^d_1,\ldots,\varphi^d_k,\varphi^d_{k+1},\ldots,\varphi^d_K\}$, where $\varphi^d_k\leq\pi\leq\varphi^d_{k+1}$. Then, the optimal discrete phase can be searched in the following way.

According to the piecewise monotonicity of the objective function, $\varphi^d_1$ is optimal in the candidate phase set $\{\varphi^d_1,\ldots,\varphi^d_k\}$, and $\varphi^d_K$ is optimal in the set $\{\varphi^d_{k+1 },\ldots,\varphi^d_K$\}. Therefore, the optimal discrete phase candidates are $\{-\varphi_1+\varphi^d_1,\varphi_1+\varphi^d_K\}$. Owing to the periodicity of the objective function, $\varphi^d_K$ and $\varphi^d_1$ are equivalent to the closest discrete phases at the left/right side of the optimal continuous phase $\varphi^*_A=-\varphi_1$. Furthermore, let us denote $(\varphi)^+$ and $(\varphi)^-$ as the nearest available discrete phase around $\varphi$. It holds true that the optimum discrete phase control $\varphi^*_A\in \{(-\varphi_1)^+, (-\varphi_1)^-\}$. Therefore, \emph{{Theorem $3$}} has been proved. \hfill\IEEEQEDclosed

%\vspace{5mm}
\balance
\bibliographystyle{IEEEtran}

\bibliography{v4}

\vfill

\end{document}